\documentclass[aps,prd,reprint,preprintnumbers,floats,epsfig,nofootinbib,amssymb,twocolumn,====
nofootinbib]{revtex4}

\usepackage{slashed}
\usepackage{comment}
\usepackage{graphicx,color}
\usepackage{epsfig}
\usepackage{subfigure}
\usepackage{epsfig}
\usepackage{epstopdf}
\usepackage{dcolumn}
\usepackage{bm}
\usepackage{color}
\usepackage{ulem}
\usepackage{enumitem}
\usepackage[colorlinks,
linkcolor=black,
anchorcolor=black,
citecolor=black
]{hyperref}

\begin{document}

\baselineskip=15pt


\title{Widening the  $U(1)_{L_\mu - L_\tau}$  $Z^\prime$ mass range for resolving the muon $g-2$ anomaly}

\author{Yu Cheng$^{1,2}$\footnote{chengyu@sjtu.edu.cn},
Xiao-Gang He$^{1,2,3}$\footnote{hexg@phys.ntu.edu.tw},
Jin Sun$^{1,2}$\footnote{019072910096@sjtu.edu.cn}}

\affiliation{${}^{1}$Tsung-Dao Lee Institute, Shanghai Jiao Tong University, Shanghai 201210, China}
\affiliation{${}^{2}$Shanghai Key Laboratory for Particle Physics and Cosmology, Key Laboratory for Particle Astrophysics and Cosmology (MOE), School of Physics and Astronomy, Shanghai Jiao Tong University, Shanghai 200240, China}
\affiliation{${}^{3}$NCTS and Department of Physics, National Taiwan University, Taipei 10617, Taiwan}

\begin{abstract}
Exchanging a $Z^\prime$ gauge boson is a favored mechanism to solve the muon $(g-2)_\mu$ anomaly.  Among such models the $Z^\prime$
from  $U(1)_{L_\mu - L_\tau}$ gauge group has been extensively studied. In this model the same interaction addressing $(g-2)_\mu$,   leads to an enhanced muon neutrino trident  (MNT) process  $\nu_\mu N \to \nu_\mu \mu \bar \mu N$ constraining the $Z^\prime$
mass to be less than a few hundred MeV.  Many other $Z^\prime$ models face the same problem.  It has long been realized that the coupling of $Z^\prime$ in the model can admit $(\bar \mu \gamma^\mu \tau + \bar \nu_\mu \gamma^\mu L \nu_\tau)Z^\prime_\mu$ interaction which does not contribute to the MNT process. It  can solve $(g-2)_\mu$  anomaly for a much wider $Z^\prime$ mass range. However this new interaction induces $\tau \to \mu \bar\nu_\mu \nu_\tau$ which rules out it as a solution to $(g-2)_\mu$ anomaly.   Here  we propose a mechanism by introducing type-II seesaw $SU(2)_L$ triplet  scalars to evade constraints from  all known data to allow a wide $Z^\prime$ mass range to solve the $(g-2)_\mu$ anomaly. This  mechanism opens a new window for $Z^\prime$ physics.
\end{abstract}

\maketitle

\noindent

{
The muon $g-2$ Collaboration at Fermilab reported their new results from Run 1 measurement of the muon
anomalous magnetic dipole moment $a_\mu$ recently~\cite{Fermilab}. Combining previous data from BNL~\cite{previous-data}, the discrepancy between
experimental data $a^{exp}_\mu$ and standard model (SM) prediction $a^{SM}_\mu$~\cite{SM-number} reinforces muon $g-2$ anomaly  and raises the confidence level  from 3.7$\sigma$ to 4.2$\sigma$~\cite{Fermilab} with,
$\Delta a_\mu = a^{exp}_\mu - a^{SM}_\mu = (251\pm 59)\times 10^{-11}$.  More precise SM calculations and experimental measurements are needed to further confirm this anomaly\footnote{Recent lattice calculation in fact favors experimental value compared with previous calculations~\cite{nature-g-2}.}.   Many theoretical ideas trying to explain this anomaly have been proposed. Exchanging a $Z^\prime$ boson from a new beyond SM $U(1)$ gauge group is one of the mostly studied mechanisms~\cite{baek-he-ko, Ma:2001md,  Altmannshofer:2014pba, Altmannshofer:2016oaq, gauge1,gauge2, gauge3, Cen:2021iwv, Kang:2021jmi, Bause:2021prv, Isidori:2021gqe}.  The $Z^\prime$ from $U(1)_{L_\mu - L_\tau}$ gauge group~\cite{lmu-ltau1, lmu-ltau2, Foot:1994vd} to solve this anomaly~\cite{baek-he-ko, Altmannshofer:2014pba, Altmannshofer:2016oaq} is among some of the favorable models.
The same $Z^\prime$ interaction with muons will modify the muon neutrino trident  (MNT) process $\nu_\mu N \to \nu_i \mu \bar \mu N$ which constrains the $Z^\prime$ mass to be below about 300 MeV or so if solution to $(g-2)_\mu$ is required~\cite{Altmannshofer:2014pba}.  Similar constraint exists for many other $Z^\prime$ models. If the $Z^\prime$ mass turns out to be larger than a few hundred MeV, are the $U(1)_{L_\mu - L_\tau}$ and other related $Z^\prime$ models completely ruled out as a possible solution to the muon $(g-2)_\mu$ anomaly? The answer is no.   In this letter, we propose a mechanism to evade the MNT constraint and to open the large $Z^\prime$ mass window to address the $(g-2)_\mu$ anomaly. We work with the $U(1)_{L_\mu - L_\tau}$ $Z^\prime$ to show that it is indeed possible to achieve this.  We find that type-II seesaw triplet $\Delta: (3, 1)$ scalars play the important role of solving the problem to widen the allowed $U(1)_{L_\mu-L_\tau}$ $Z^\prime$ mass range.

In the simplest $U(1)_{L_\mu - L_\tau}$ model, the left-handed $SU(3)_C\times SU(2)_L\times U(1)_Y$ doublets $L_{L\;i}: (1, 2, -1/2)$ and the right-handed singlets $e_{R\;i}: (1,1,-1)$ transform under the gauged $U(1)_{L_\mu-L_\tau}$ group as $0,\;1,\;-1$ for the first, second and third generations, respectively. The $Z^\prime$ gauge boson of the model only interacts with leptons in the  weak interaction basis~\cite{lmu-ltau1, lmu-ltau2}
\begin{eqnarray}
- \tilde g (\bar \mu \gamma^\mu \mu - \bar \tau  \gamma^\mu \tau + \bar \nu_\mu \gamma^\mu L \nu_\mu - \bar \nu_\tau \gamma^\mu L \nu_\tau) Z^\prime_\mu \;, \label{zprime-current}
\end{eqnarray}
where $L(R) = (1 - (+)\gamma_5)/2$.
Exchanging  $Z^\prime$ at one loop level can generate muon $g-2$~\cite{Leveille:1977rc}
\begin{eqnarray}
\Delta a_\mu^{Z^\prime}= {\tilde g^2\over 8\pi^2} {m^2_\mu \over m^2_{Z^\prime}} \int^1_0 {2 x^2(1-x) dx \over 1-x + (m^2_\mu/m^2_{Z^\prime}) x^2}\;. \label{leading}
\end{eqnarray}
In the limit $m_{Z^\prime} >> m_\mu$, $\Delta a^{Z^\prime}_\mu = (\tilde g^2/ 12 \pi^2)(m^2_\mu/m^2_{Z^\prime} )$.

The above $Z^\prime$ interaction will also modify the cross section for the MNT process compared with the SM prediction.
Several experiments give the measurements for the ratio $\sigma_{exp}/\sigma_{SM}|_{trident}$: $1.58\pm 0.57$, $0.82\pm 0.28$ and $0.72^{+1.73}_{-0.72}$ from CHARM-II~\cite{CHARM-II:1990dvf}, CCFR~\cite{CCFR:1991lpl} and NuTeV~\cite{NuTeV:1999wlw}, respectively.
Using the most accurate CCFR data, the allowed region in $\tilde g-m_{Z^\prime}$ plane for solving the $(g-2)_\mu$ anomaly has been obtained in Ref.~\cite{Cen:2021iwv,Altmannshofer:2014pba}. Their analysis show that the $Z^\prime$ mass is constrained to be less than a few hundred MeV ($\sim$ 300 MeV).
The details for conflict are shown as the follows. For a large $m_{Z'}$ with GeV scale, one has
\begin{eqnarray}
{\sigma_{Z^\prime}\over \sigma_{SM}} \vert_{trident}= { (1+4s_W^2 + 8 \tilde g^2 m^2_W/g^2 m^2_{Z^\prime})^2 + 1 \over 1 + (1+4 s^2_W)^2}\;, \label{trident-neutrino}
\end{eqnarray}
where $g$ is the $SU(2)_L$ coupling constant and $s_W = \sin\theta_W$ with $\theta_W$ being the Weinberg weak mixing angle.

For $m_{Z^\prime} >> m_\mu$, assuming the difference $\Delta a_\mu$ is due to $\Delta a^{Z^\prime}_\mu$, we have
${\tilde g^2/m^2_{Z'}} = (2.66\pm 0.63)\times 10^{-5} \mbox{GeV}^{-2}$. Inserting this into Eq.(\ref{trident-neutrino}), we obtain
$\sigma_{Z^\prime}/ \sigma_{SM} = 5.86$ which is several times larger than experimentally allowed upper bounds~\cite{CHARM-II:1990dvf, CCFR:1991lpl, NuTeV:1999wlw}.  In this context, $Z^\prime$ contribution to MNT  process should be lowered to a factor 20 to be within the one sigma range of the weighted average $0.95\pm 0.25$ from the above three data.

If there is a mixing $\theta$ between $\mu$ and $\tau$, the MNT process will be suppressed by a factor of $\theta^2$, and at the same time the parameter ${\tilde g^2/m^2_{Z'}}$ is suppressed by the factor $\theta^2 m_\tau/m_\mu$. This  can reduce the MNT constraint. However, with both flavor diagonal and off-diagonal $Z^{\prime}$ couplings to $\mu$ and $\tau$, the lepton flavor violating (LFV) processes,  such as $\tau \rightarrow \mu \gamma$ and $\tau \rightarrow 3 \mu$ will be  induced. These processes put very strong constraints making it impossible for the model to explain the $(g-2)_\mu$ anomaly.
Therefore, it is crucial to have a mechanism to widen the $Z^\prime$ mass range with larger than 300 MeV by forbidding flavor changing processes and to open a new window of searching for $Z^\prime$ physics, such as at a future collider.

To this end, we notice that a $U(1)_{L_\mu-L_\tau}$ model proposed long times ago includes totally off-diagonal $Z^\prime$-lepton interactions in the mass eigen-states~\cite{Foot:1994vd}
\begin{eqnarray}
- \tilde g (\bar \mu \gamma^\mu \tau +  \bar \tau  \gamma^\mu \mu + \bar \nu_\mu \gamma^\mu L \nu_\tau  + \bar \nu_\tau \gamma^\mu L \nu_\mu) Z^\prime_\mu \;. \label{zprime-current}
\end{eqnarray}
The above interaction does not contribute to MNT and $\tau\to 3\mu, \mu \gamma$ processes so that $Z^\prime$ mass is not constrained while  solving $(g-2)_\mu$ anomaly. In the limit $m_{Z'} >> m_{\tau,\mu}$, the new contribution to $(g-2)_\mu$ is given by~\cite{baek-he-ko}
\begin{eqnarray}
\Delta a^{Z^\prime}_\mu = {\tilde g^{2} m^2_\mu\over 12 \pi^{2}m_{Z^{\prime}}^{2}}\left({3 m_{\tau}\over m_{\mu}}-2\right)\;.
\end{eqnarray}
We see that there is an enhancement factor of $(3 m_\tau/m_\mu-2)$ to $\Delta a_\mu$. In this case  $\tilde g^2/m^2_{Z^\prime}$ is about 48 times smaller than the previous case
which correspondingly makes  $m_{Z^\prime}$ below a few hundred MeV no longer.

However, the $Z^\prime$ interaction  will induce $\tau \to \mu \bar \nu_\mu \nu_\tau + \mu \bar \nu_\tau \nu_\mu$. Including the SM $W$ and the new $Z^\prime$ contributions, we have the effective Lagrangian $- L_{W+ Z^\prime}$ as
\begin{eqnarray}\label{SM}
{g^2\over 2 m^2_{W}}  \bar \nu_\tau \gamma^\mu L \nu_\mu \bar \mu \gamma_\mu L \tau  + {\tilde g^2\over m^2_{Z^\prime}}(\bar \nu_\tau \gamma^\mu L \nu_\mu + \bar \nu_\mu \gamma^\mu L \nu_\tau)\bar \mu \gamma_\mu  \tau\;.\nonumber\\
\end{eqnarray}
Data from $R_{\tau\mu}=\Gamma(\tau \to \mu \nu \bar \nu)/\Gamma_{SM}(\tau \to \mu \nu \bar \nu)=1 .0066\pm0.0041$~\cite{Altmannshofer:2016brv}, provides a very stringent constraint for this model. In fact  experimental value $R_{\tau\mu}$ excludes this model, if the model is required to solve the $(g-2)_\mu$ anomaly simultaneously, by more than   5$\sigma$ level.
Thus, one needs to
reduce the $Z'$ contribution to $\tau \to \mu \bar \nu \nu$ to make it satisfy this experimental constraint while addressing $(g-2)_\mu$ anomaly.

We find  that introducing type-II seesaw triplet scalars $\Delta_i: (1,3,1)$~\cite{Lazarides:1980nt,Mohapatra:1980yp,Konetschny:1977bn,Cheng:1980qt,Magg:1980ut,Schechter:1980gr}, which provides small neutrino masses, can solve the problem.
The component field $\Delta$ is given by
\begin{eqnarray}
\Delta = \left ( \begin{array}{cc}
\Delta^+/\sqrt{2}&\;\;\Delta^{++}\\
\Delta^0&\;\; - \Delta^+/ \sqrt{2}
\end{array}
\right )\;.
\end{eqnarray}
We now provide some details of the model. We firstly discuss how to obtain the $Z^\prime$ interaction in Eq.(\ref{zprime-current}). For this purpose,  one needs to introduce three Higgs doublets $H_{1,2,3}: (1,2, 1/2)$ ($<H_i> = v_{i}/\sqrt{2}$) with   $U(1)_{L_\mu - L_\tau}$ charges ($0, 2, -2$)
and to impose a unbroken exchange symmetry $Z^\prime \to - Z^\prime$,  $H_1 \leftrightarrow H_1$ and $H_2 \leftrightarrow H_3$ with $v_2=v_3=v$,  while the other SM fields do not transform.  In this case the $Z^\prime$ interaction and Yukawa terms to leptons are given by
\begin{eqnarray}
L _{H}= &&- \tilde g (\bar l_2 \gamma^\mu L l_2- \bar l_3  \gamma^\mu L l_3 + \bar e_2 \gamma^\mu R e_2 - \bar e_3 \gamma^\mu R e_3) Z^\prime_\mu\nonumber\\
&&- [Y^l_{11} \bar l_1 R e_1 + Y^l_{22} (\bar l_2 R e_2 +\bar l_3 R e_3 ) ] H_1 \nonumber\\
&&-Y^l_{23} (\bar l_2 R e_3 H_2 +\bar l_3 R e_2 H_3 ) + H.C.
\end{eqnarray}

The transformation between  the charged lepton mass eigen-state and weak eigen-state basis is given by
\begin{eqnarray}\label{eigen}
\left (
\begin{array}{c}
\mu\\
\tau
\end{array}
\right )
= {1\over \sqrt{2}}
\left (
\begin{array}{rr}
1&\;-1\\
1&\;1
\end{array}
\right )
\left (\begin{array}{c}
e_2\\
e_3
\end{array}
\right ),\;
\end{eqnarray}
The neutrino can conduct the above same transformation.
These lead to $Z^\prime$ interaction of the form in Eq.(\ref{zprime-current}).

Note that in the above analysis there is no kinetic mixing between the $U(1)_Y$ and $U(1)_{L_\mu-L_\tau}$ gauge fields.
Although $H_{2,3}$ carry both the $U(1)_Y$ and $U(1)_{L_\mu-L_\tau}$ charges, there may exist mixing between the $U(1)_Y$ gauge boson $Y_\mu$ and $Z^\prime_\mu$. However, we find this does not happen since the requirement of unbroken exchange symmetry in the model. 
The exchange symmetry may be broken by quantum corrections. 
But we have check carefully that at one loop level there is no such broken terms due to the same coupling of $H_2$ and $H_3$ ($\Delta_2$ and $\Delta_3$) to the other fields if the vevs of $H_2$($\Delta_2$) and $H_3$($\Delta_3$) have the same value. For example we find no $Z-Z'$ mixing at tree level which may result in breaking of the exchange symmetry, due to the cancellation of the opposite $U(1)_{L_\mu-L_\tau}$ charges for $H_2$ and $H_3$ ($\Delta_2$ and $\Delta_3$) , and also the Higgs potential at one loop level does not change the equality of the vevs. So, the breaking should happen at higher order two-loop level in principle, which is
very small.
The $Z^\prime$ will receive a mass $4\tilde g^2 (v^2_2+v^2_3) = 8 \tilde g^2 v^2$
 with the vev $v^2$ below $4m^2_W/g^2$ implying $Z^\prime$ mass cannot be too large.
To lift up $m_{Z'}$,
one can introduce a singlet scalar $S: (1,\;1,\;0)(1)$ with a non-zero vev $v_s/\sqrt{2}$ to obtain $m^2_{Z^\prime}= \tilde g^2( v_s^2 + 8 v^2)$. Therefore $Z^\prime$ can have an arbitrarily  mass depending on the values of $v$ and $v_s$. At the electroweak scale $U(1)_{L_\mu-L_\tau}$ $Z^\prime$ has been shown to be allowed by experimental data~\cite{Heeck:2011wj}.

We now explain how to reduce $Z^\prime$ contribution to $\tau \to \mu \nu \bar \nu$ using the type-II seesaw triplet scalars.
For this purpose we add three triplets $\Delta_{1,2,3}$ ($<\Delta_i> = \tilde v_i/\sqrt{2}$) with $U(1)_{L_\mu - L_\tau}$ charges ($0, -2, 2$).
Under the exchange symmetry discussed above, $\Delta_1 \leftrightarrow \Delta_1$, $\Delta_2 \leftrightarrow \Delta_3$, we have the following Yukawa terms
\begin{eqnarray}
L _\Delta = &&- [Y^\nu_{11} \bar l^c_1 L l_1  + Y^\nu_{23} (\bar l^c_2 L l_3 + \bar l^c_3 L l_2 )]\Delta_1  \nonumber \\
&&-Y^\nu_{22} (\bar l^c_2 L l_2 \Delta_2 +\bar l^c_3 L l_3 \Delta_3 )] + H.C.\;,
\end{eqnarray}
 The unbroken exchange symmetry imposed would imply $\tilde v_2 = \tilde v_3$.
The non-zero $\tilde v_i$ not only generate neutrino masses, but also modify the  precision electroweak parameter $\rho$~\cite{Zyla:2020zbs} to deviate from one. This results in restricting $\tilde v_i$  so small that their effects on $W$ mass can be safely neglected.

In the basis shown in Eq.(\ref{eigen}), we have
\begin{eqnarray}
L_\Delta = -&& \left [\bar l^c_\mu L l_\mu (Y^\nu_{22}( \Delta_2 +\Delta_3) - 2 Y^\nu_{23} \Delta_1) \right. \nonumber \\
&&\left. + \bar l^c_\tau L  l_\tau (Y^\nu_{22} (\Delta_2 +\Delta_3)  + 2 Y^\nu_{23} \Delta_1)\right. \nonumber\\
&& \left.+ 2\bar l^c_\mu L  l_\tau (Y^\nu_{22}( \Delta_2 -\Delta_3))
\right]/2 + H.C.\;.
\end{eqnarray}
Expanding out the above interaction in terms of the component fields $\Delta^{0, +, ++}$, $\bar l^c_i L l_j \Delta$ can be written as
\begin{eqnarray}
 \bar\nu^c_i L \nu_j \Delta^0  - \bar e^c_i L e_j \Delta^{++} - {1\over \sqrt{2}}(\bar \nu^c_i L e_j+ \bar e^c_i L \nu_j )\Delta^+.
\end{eqnarray}
This form also applies to $l_1$.  Thus, $L_\Delta$  in terms of the component fields is given by
\begin{eqnarray}
&&
\left (\sqrt{2} (\bar \nu_e^c, \bar \nu_\mu^c, \bar \nu_\tau^c ) M(\Delta^+)  + (\bar e^c, \bar \mu^c, \bar \tau^c ) M(\Delta^{++})\right ) L
(e,\mu,\tau)^T
\nonumber\\
&&- (\bar \nu_e^c, \bar \nu_\mu^c, \bar \nu_\tau^c ) M(\Delta^0)  L
(\nu_e,\nu_\mu,\nu_\tau)^T
\;,
\end{eqnarray}
where the non-zero entries 
of $M(\Delta)$ are
\begin{eqnarray}\label{Matrix}
&&M_{11}=Y^\nu_{11} \Delta_1, \;
M_{23}= M_{32}=   {1\over 2} Y^\nu_{22}(\Delta_2-\Delta_3),\;\nonumber\\
&&M_{22,33} = {1\over 2} Y^\nu_{22}(\Delta_2 +\Delta_3) \mp Y^\nu_{23}\Delta_1\;.
\end{eqnarray}
Exchanging $\Delta_i^{+,++}$ will generate a non-zero $\Delta a_\mu^\Delta$ given by~\cite{Leveille:1977rc}. In the limit $m_\Delta >> m_\tau$, the contribution is
\begin{eqnarray}
- {m^2_\mu \over 16 \pi^2} \left [  \left({|Y^\nu_{22}|^2\over m^2_{\Delta_2}} + {|Y^\nu_{22}|^2\over m^2_{\Delta_3}}\right) +  {|Y^\nu_{23}|^2\over m^2_{\Delta_1}} \right ]\;.
\end{eqnarray}

Exchanging $\Delta^+_i$ at tree level will contribute to $\tau \to \mu \bar \nu \nu$, the $M(\tau \to \mu \bar \nu \nu)$ is given by
\begin{eqnarray}
&&\rho_{\mu\mu}(\bar \nu_\mu\gamma^\mu L \nu_\mu + \bar \nu_\tau \gamma^\mu L\nu_\tau)\bar \mu\gamma_\mu L \tau  \nonumber\\
&&+(\rho_{\tau\mu}\bar \nu_\tau \gamma^\mu L\nu_\mu
+ \rho_{\mu\tau} \bar \nu_\mu \gamma^\mu L\nu_\tau) \bar \mu\gamma_\mu L \tau
\end{eqnarray}
where $\rho_{\mu\mu,\tau\mu}= \left({|Y^\nu_{22}|^2/ m^2_{\Delta_2}} \mp {|Y^\nu_{22}|^2/ m^2_{\Delta_3}}\right)/4$  and $\rho_{\mu\tau}=\rho_{\tau\mu}- {|Y^\nu_{23}|^2/ m^2_{\Delta_1}}$.

By taking into account the contribution $L_{W+Z'}$ in Eq.(\ref{SM}), one can obtain the total amplitude $M_{total}(\tau \to \mu \nu \bar \nu)$ as
\begin{eqnarray}
\label{Mtaudecay}
&& \left({\rho_{\tau\mu} \over 2} - {g^2\over 4m^2_W} - {\tilde g^2\over m^2_{Z^\prime}} \right)\bar \mu \gamma_\mu \tau \bar \nu_\tau \gamma^\mu L \nu_\mu \nonumber\\
&&-\left( {\rho_{\tau\mu} \over 2} -{g^2\over 4m^2_W}\right) \bar \mu \gamma_\mu \gamma_5 \tau \bar \nu_\tau \gamma^\mu L \nu_\mu \nonumber\\
&&+ \left[ \left({\rho_{\mu\tau} \over 2} - {\tilde g^2\over m^2_{Z^\prime}} \right) \bar \mu \gamma_\mu \tau
- \left( {\rho_{\mu\tau} \over 2} \right)  \bar \mu \gamma_\mu\gamma_5 \tau \right] \bar\nu_\mu \gamma^\mu L\nu_\tau\nonumber\\
&&+( \rho_{\mu\mu})\bar \mu \gamma_\mu L \tau(\bar \nu_\mu\gamma^\mu L \nu_\mu + \bar \nu_\tau \gamma^\mu L\nu_\tau) \;.  \label{totaltau}
\end{eqnarray}

From Eq.(\ref{Matrix}) one can easily see $\Delta^{++}$ will also generate $\tau \to \mu  \mu \bar \mu$ at tree level  with the branching ratio as
\begin{equation}
B r(\tau \rightarrow 3 \mu)=\frac{m_{\tau}^{5} \tau_{\tau}}{3 (16\pi)^{3}} \left({|Y^\nu_{22}|^2\over m^2_{\Delta_2} }- {|Y^\nu_{22}|^2\over m^2_{\Delta_3}}\right)^2.
\end{equation}
We must check if the experimental data $B r^{\exp }(\tau \rightarrow 3 \mu)<2.1 \times 10^{-8}$ in the 90\% C.L.~\cite{Zyla:2020zbs} already rule out the model. We find that as long as $|Y^\nu_{22}|^2(1/m^2_{\Delta_2} - 1/ m^2_{\Delta_3})<3 \times 10^{-8}$,  the constraint from data can be satisfied. If $\Delta_2$ and $\Delta_3$ are degenerate in mass, $\rho_{\mu\mu} =0$.   This is actually consistent with the unbroken exchange symmetry required.
Therefore we take the degenerate case in our numerical analysis. In this case the process $\tau \to \mu\mu \bar \mu$ will vanish. This also eliminates potential problem from  one loop induced $\tau \to \mu \gamma$.

Note that in the model $\Delta_1$ contributes to $\mu \to e\nu\bar \nu$~\cite{Pich:2013lsa} and therefore affects data for $R_{\mu e}^{SM} = 0.972559 \pm 0.000005$ and $R_{\mu e}^{PDG}=0.979 \pm 0.004$~\cite{Zyla:2020zbs}.
To ensure $\mu\to  e \nu\bar \nu$ is not affected while solving the $\tau \to \mu \nu \bar \nu$ decay $R_{\tau\mu}$ problem, we can take $Y^\nu_{11,23}<<Y^\nu_{22}$.
Under this assumption, the process $\tau \to e \nu \bar\nu$ can also be neglected.

Let us come back to constraints from $(g-2)_\mu$ anomaly and $\tau \to \mu \nu \bar \nu$ decay. From Eq.(\ref{totaltau}) we find that a large portion of new contributions to $\tau \to \mu \nu \bar \nu$ can be cancelled out by taking $\rho_{\tau\mu} /2 - \tilde g^2/m^2_{Z^\prime} = 0$.
Since we work with the degenerate case $\rho_{\mu\mu}=0$, it  further reduces new contributions to the process. The results are shown in Fig.\ref{g2taudecay}. We find that while ${\tilde g^2 / m^2_{Z^\prime}} \in (3,8) \times 10^{-7}$  GeV$^{-2}$ and
$|Y^\nu_{22}|^2 /m^2_{\Delta} \in (0.07,2) \times 10^{-6}$ GeV$^{-2}$, constraints from both the muon $(g-2)_\mu$ and $R_{\tau\mu}$ are satisfied. One can obtain the numerical values with the order of $10^{-2}$ for $\tilde g^2$  and a few hundred GeV for $m_{Z^\prime}$, and similarly a few $ 10^{-2}$ for $Y^\nu_{22}$ and a few hundred GeV to  TeV for $m_\Delta$.
The Belle-II sensitivity to probe the light $Z'$ with diagonal interactions has been discussed recently~\cite{Jho:2019cxq,Belle:2021feg}. However, their analysis can not apply to the large $m_{Z^{\prime}}$ region with hundreds GeV in our model. In order to analyze the potentiality of probing our model in Belle-II experiment, we can further extend to off-diagonal interactions with $\tilde g=2\times 10^{-2}$ and $m_{Z'}=6$GeV, which results in $\tilde g^2/m^2_{Z'}=10^{-5}\mbox{GeV}^{-2}$ above the allowed region. However, we expect that in the near future Belle-II can probe the off-diagonal interactions in our model. 
\begin{figure}[!t]
	\centering
	\includegraphics[width=0.4\textwidth]{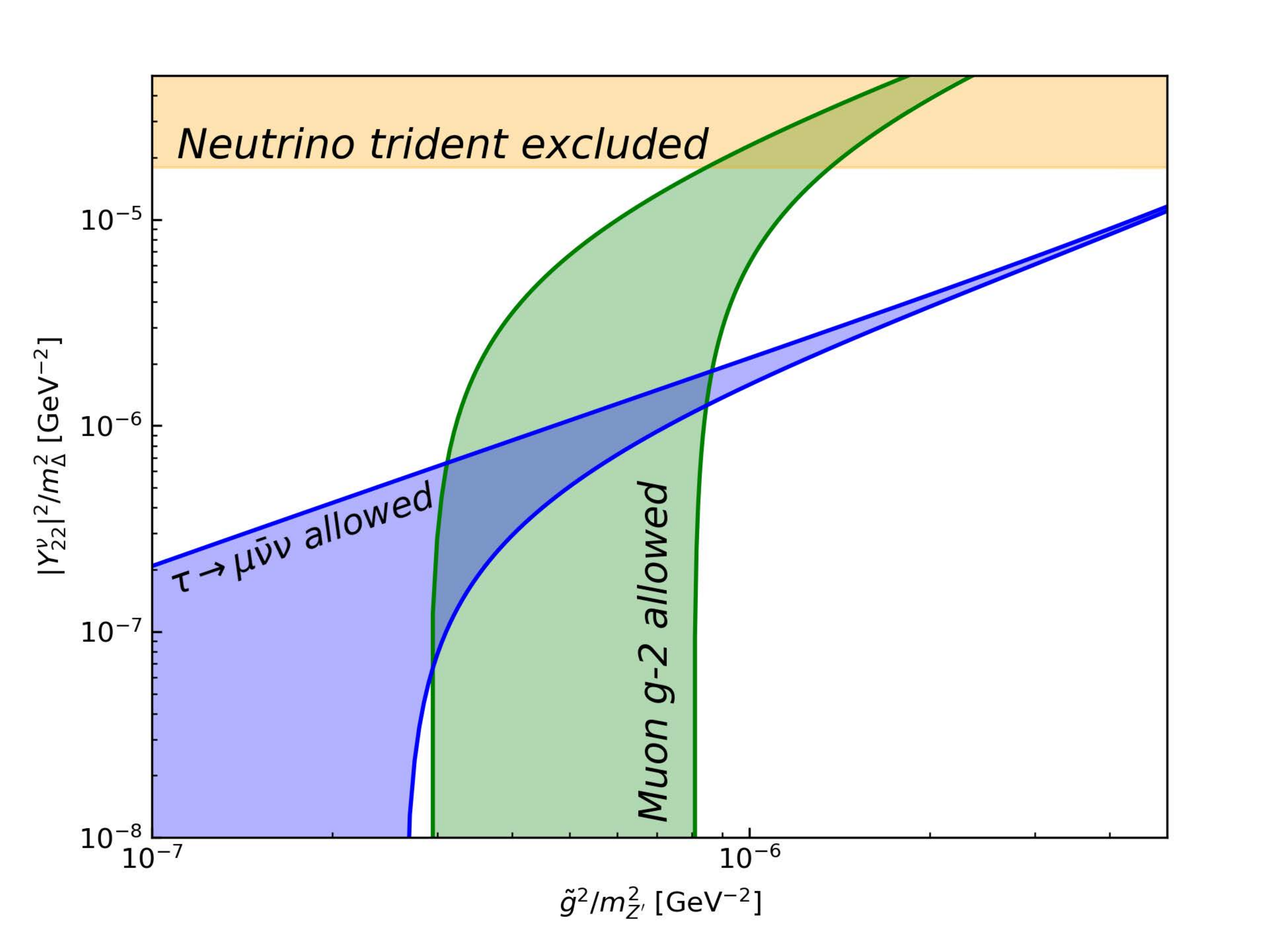}
	\caption{ The allowed parameter space in the $\tilde g^2/m^2_{Z'}-|Y_{22}^\nu|^2/m^2_{\Delta}$  plane. The shaded areas for explaining the muon $g-2$ and  satisfying $\tau \rightarrow \mu \bar{\nu} \nu$ constraint are within 2 $\sigma$ errors. The  shaded area is excluded by MNT process at 95\% confidence level. }
	\label{g2taudecay}
\end{figure}

Exchanging  $\Delta^+$ at tree level will regenerate contributions to MNT  process with ${\sigma_\Delta / \sigma_{SM}}|_{trident}$ as
\begin{eqnarray}
&& \left[(1+4 s^2_W- (m^2_W/g^2)\delta_{\nu_\mu \mu})^2+ (1-(m^2_W/g^2)\delta_{\nu_\mu\mu})^2 \right.\nonumber\\
&&\left.+(-(m^2_W/g^2)\delta_{\nu_\tau \mu})^2 \right]/[(1+4 s^2_W)^2 + 1],
\end{eqnarray}
where $\delta_{\nu_\mu \mu} = {|Y^\nu_{22}|^2/ m^2_{\Delta_2}} + {|Y^\nu_{22}|^2/ m^2_{\Delta_3}} + 4 {|Y^\nu_{23}|^2 / m^2_{\Delta_1}}$ and $\delta_{\nu_\tau \mu} = {|Y^\nu_{22}|^2/ m^2_{\Delta_2} }- {|Y^\nu_{22}|^2/ m^2_{\Delta_3}}$.

From the above equation, we see that the first two terms are to lower the SM contribution, and the third term is zero in the degenerate case. Therefore $(\sigma_\Delta/\sigma_{SM})|_{trident} <1$,
numerically $(\sigma_\Delta /\sigma_{SM})|_{trident}$ is predicted to be in the range $0.926 \sim 0.997$, which is within 1$\sigma$ error of the weighted average data $0.95\pm 0.25$.

The $Z^{\prime}$ and triplet scalar will additionally contribute to the process $e^{+} e^{-} \rightarrow Z \rightarrow \mu^{+} \mu^{-}/\tau^{+} \tau^{-}/\nu \bar{\nu}$ via one-loop correction to the $Z-ff$ vertex. The contribution is proportional to the coupling $\tilde{g}^2/m^2_{Z^{\prime}}$ and $|Y^\nu_{22}|^2 /m^2_{\Delta}$. Only considering the $Z^{\prime}$ contribution as studied in Ref.~\cite{Altmannshofer:2016brv}, the process constrains the coupling $\tilde{g}^2/m^2_{Z^{\prime}}<(10^{-5} - 10^{-4})$ GeV$^{-2}$,  which is 10-100 times larger than the allowed region to explain muon $(g-2)_\mu$ result. In our work, although there are contribution from triplet scalar, but the coupling $\tilde{g}^2/m^2_{Z^{\prime}}$ and $|Y^\nu_{22}|^2 /m^2_{\Delta}$ is at the order of $(10^{-7} - 10^{-6})$ GeV$^{-2}$ so that the constraints can be safely satisfied.

Note that the interaction in Eq.(\ref{zprime-current}) can produce distinctive signature of searching for $Z^\prime$ effects at $e^+e^-$ and $pp$ colliders
through intermediate state $\gamma^*$ and $Z^*$, $\gamma^*/Z^* \to \mu^\pm \tau^\mp + (Z^\prime \to \mu^- \tau^+, \mu^+ \tau^-)$.
The final states $\mu^\pm \mu^\pm \tau^\mp \tau^\mp$ would serve as unique signature for the model~\cite{Altmannshofer:2016brv}, which can also be probed at a future $Z$ factory.
There is another source for such processes coming from $e^+e^- (pp) \to \Delta^{++}\Delta^{--} \to \mu^\pm \mu^\pm \tau^\mp \tau^\mp$. However, one can easily discriminate  this two kinds of different signatures sources by studying  the  structure for the $\mu^\pm \tau^\mp$ resonance from $Z'$ or $\mu^\pm \mu^\pm$ and $\tau^\mp \tau^\mp$ from $\Delta$, respectively.
We urge our experimental colleagues to carry out such analysis for different $Z^\prime$ masses. More  phenomenological studies  of the model and a class of related models  will be discussed elsewhere.

In conclusion, we have proposed a mechanism by introducing type-II seesaw $SU(2)_L$ triplet  scalars to evade constraints from  MNT process, $\tau \to \mu \bar\nu_\mu \nu_\tau$ and several other processes while solving  $(g-2)_\mu$ anomaly by exchanging $Z^\prime$ from the $U(1)_{L_\mu - L_\tau}$  model.   The allowed $Z^\prime$ mass range is widened,  which opens a new window for $Z^\prime$ physics.

\section*{Acknowledgments}
This work was supported in part by Key Laboratory for Particle Physics, Astrophysics and Cosmology, Ministry of Education, and Shanghai Key Laboratory for Particle Physics and Cosmology (Grant No. 15DZ2272100), and in part by the NSFC (Grant Nos. 11735010, 11975149, and 12090064). XGH was supported in part by the MOST (Grant No. MOST 106-2112-M-002-003-MY3 ).

\end{document}